\documentclass[12pt]{article}

\usepackage{amssymb,amsmath}
\usepackage{graphbox}
\usepackage{graphicx}
\usepackage{slashed}
\begin{document}

\begin{titlepage}

\begin{flushright}
arXiv:2509.25488
\end{flushright}
\vskip 2.5cm

\begin{center}
{\Large \bf Renormalization Group Running of the Parity\\
Operator in Lorentz-Violating Quantum Field Theory}
\end{center}

\vspace{1ex}

\begin{center}
{\large Brett Altschul\footnote{{\tt altschul@mailbox.sc.edu}}}

\vspace{5mm}
{\sl Department of Physics and Astronomy} \\
{\sl University of South Carolina} \\
{\sl Columbia, SC 29208}
\end{center}

\vspace{2.5ex}

\medskip

\centerline {\bf Abstract}

\bigskip

In conventional relativistic quantum field theory, the discrete operators \textbf{C}, \textbf{P}, and
\textbf{T} are matrix operators with no renormalization scale dependence. However, in a Lorentz-violating
theory with a fermion $f^{\mu}$ term in the action, these operators may acquire nontrivial renormalization
group behavior. Because the $f^{\mu}$ term may actually be exchanged in the action for an
equivalent $c^{\nu\mu}$ term, the scale dependence depends explicitly on the renormalization scheme,
even at one-loop order. The scheme dependence means it is always possible to set the scale dependence parameter
$1-X$ to zero, but
for analyses of some high-energy electron-photon processes, using a scheme with $X=0$---and thus definite
scale dependences for \textbf{C}, \textbf{P}, and \textbf{T}---may nonetheless be more convenient.

\bigskip

\end{titlepage}

\newpage

\section{Introduction}

Special relativity underlies both of the theories that we use to describe fundamental physics: the
standard model of particle physics and the general theory of relativity. The Lorentz symmetry of
special relativity manifests itself in somewhat different ways in these two regimes, but it is a
through line connecting both of them. Given the importance of this symmetry, there has long been
significant interest in testing whether it holds exactly or merely approximately. This interest is
quite natural.
We already know that an apparent, but in reality only approximate, symmetry can have profound implications
and provide key insights about where new fundamental physics might be found.

In this millennium, thanks primarily to advances in effective field theory methods, it has been
possible to parameterize violations of rotational isotropy and Lorentz boost invariance in a
systematic way. This, in turn, has prompted experimental interest in many types of Lorentz violation
that had previously flown under scientists' radar.
These kinds of
renewed experimental searches have not found any compelling evidence that Lorentz symmetry is not
exact, but they remain ongoing.

The local effective field theory used for describing Lorentz violation with known standard model fields
is prosaically known as the standard model extension (SME)~\cite{ref-kost1,ref-kost2}.
Because of the close connection between Lorentz invariance and \textbf{CPT}
invariance~\cite{ref-greenberg}, the SME is also capable of describing generically
any forms of \textbf{CPT} violation that are stable, unitary, local, and constructed out of the standard model's
fermion and boson fields. (Gravity is slightly trickier.)
The SME (or at least the particle theory part) is a quantum field theory (QFT), and properly understanding
a QFT always entails understanding the characteristics of its $\mathcal{O}(\hbar)$
and higher loop corrections.
The minimal SME constitutes the subtheory of the full SME framework that should be
expected to be renormalizable, as it contains only the finite number of local, Hermitian, and
gauge-invariant operators that can be constructed up to mass dimension no greater than four.
The minimal SME action looks much like the ordinary standard model action, except that the field
operators it contains are not constrained to have all their Lorentz indices fully contracted.
This minimal theory is often the most logical
framework for analyzing the results of experimental Lorentz and \textbf{CPT} tests.
There has been a fair amount of good work on the elucidation of the renormalization
properties of the minimal SME, especially
at one-loop order~\cite{ref-kost3,ref-berr,ref-collad-3,
ref-collad-2,ref-collad-1,ref-gomes,
ref-anber,ref-ferrero3,ref-brito1}. However, our understanding of the theory at the quantum level
is by no means complete.

\section{Fermion Lorentz Violation}

The free fermion sector of the SME contains the operators that control the propagation of
spin-$\frac{1}{2}$ particles. In the minimal SME, with only superficially
renormalizable operators, the Lagrange density for a single species of fermion is
\begin{equation}
\label{eq-L}
\mathcal{L}_{\psi}=\bar{\psi}(i\Gamma^{\mu}\partial_{\mu}-M)\psi.
\end{equation}
The $\Gamma$ and $M$ terms generalize the usual Dirac Lagrange density's $\gamma$ and $m$, by
including all possible matrix structures, contracted with the fixed vector and tensor backgrounds that
describe the Lorentz violation. These background structures enter in the forms
\begin{eqnarray}
\Gamma^{\mu} & = & \gamma^{\mu}+\Gamma_{1}^{\mu}=
\gamma^{\mu}+c^{\nu\mu}\gamma_{\nu}+d^{\nu\mu}\gamma_{5}
\gamma_{\nu}+e^{\mu}+if^{\mu}\gamma_{5}+\frac{1}{2}g^{\lambda\nu\mu}
\sigma_{\lambda\nu} \\
M & = & m+im'\gamma_{5}+M_{1}=m+im'\gamma_{5}+a^{\nu}\gamma_{\nu}+b^{\nu}\gamma_{5}\gamma_{\nu}
+\frac{1}{2}H^{\mu\nu}\sigma_{\mu\nu}.
\end{eqnarray}
Some of the coefficients
in ${\cal L}$ are more interesting than others. For example, $m'$ is not Lorentz violating, but rather
represents a Majorana mass term that may be absorbed into $m$ in the noninteracting
Lorentz-invariant theory by a field redefinition.

The specific form of interaction the fermions experience is not especially important, as regards the
discrete symmetry structure we shall be discussing. For the purpose of definite calculations, we shall
use a Yukawa interaction, with scalar and interaction Lagrange density
\begin{equation}
\label{eq-Lphi}
{\cal L}_{\phi}=\frac{1}{2}(\partial^{\mu}
\phi)(\partial_{\mu}\phi)-\frac{1}{2}\mu^{2}\phi^{2}-\frac{\lambda}{4!}\phi^{4}
-\bar{\psi}G\psi\phi.
\end{equation}
However, nothing about our analysis should depend qualitatively on having this particular interaction.
Generally, calculations of Lorentz-violating fermion self-energies in a Yukawa theory and an Abelian
gauge theory have similar levels of complexity.  Each theory can have certain vertex insertions
not present in the other. In a gauge theory, the gauge symmetry requires that $\Gamma$ modify not
just free fermion propagation but also the boson-fermion vertex. In a Yukawa theory, the vertex can
be modified in a different way; $G$, like $M$, may include all possible Dirac matrix structures,
\begin{equation}
G=g+ig'\gamma_{5}+G_1=g+ig'\gamma_{5}+I^{\mu}\gamma_{\mu}+J^{\mu}\gamma_{5}\gamma_{\mu}
+\frac{1}{2}L^{\mu\nu}\sigma_{\mu\nu}.
\end{equation}
The full one-loop renormalization of the Yukawa theory including all these terms (as well as additional
Lorentz violation in the pure scalar propagation sector) was dealt with in Ref.~\cite{ref-ferrero3};
however, that level of generality is not necessary here, so we shall restrict attention to a theory
with a standard Yukawa vertex $G=g$.

Without the condition of Lorentz symmetry---which amounts to a condition that the operators in the
Lagrange density of the theory should have no free Lorentz indices---we
can see that there are many more dimension-three
and dimension-four operators than in a conventional QFT.
However, there is actually some redundancy in the SME description, and
there exists an exact equivalence between two different types of Lorentz-violating fermion theories. This
redundancy is at the root of all the observations we shall make in this paper.
There is a transformation of the SME Lagrange density with solely a $f^{\mu}$ term into one
with just a $c^{\nu\mu}$ term instead~\cite{ref-altschul8}. The theory with $f^{\mu}$, defined by
\begin{equation}
\mathcal{L}_{f}=\bar{\psi}[i(\gamma^{\mu}+if^{\mu}\gamma_{5})\partial_{\mu}-m]\psi
\end{equation}
is converted into
\begin{equation}
\mathcal{L}_{c}=\bar{\psi}'[i(\gamma^{\mu}+c^{\nu\mu}_{\mathrm{eff}}\gamma_{\nu})\partial_{\mu}-m]\psi'
\end{equation}
by a local linear transformation acting on the four-component fermion field,
\begin{equation}
\label{eq-f-trans}
\psi'=e^{\frac{i}{2}f^{\nu}\gamma_{\nu}\gamma_{5}G(-f^2)}\psi,
\end{equation}
where $G(\xi)=\frac{1}{\sqrt{\xi}}\tan^{-1}\sqrt{\xi}$. The $c^{\nu\mu}$ tensor in the transformed
Lagrange density is
\begin{equation}
\label{eq-ceff-old}
c^{\nu\mu}_{\mathrm{eff}}=\frac{f^{\mu}f^{\nu}}{f^2}\left(\sqrt{1-f^2}-1\right)
\approx-\frac{1}{2}f^{\nu}f^{\mu},
\end{equation}
with the approximate form applying when $|f^{2}|$ is small. We shall work in this approximate regime, since
physical Lorentz violation, if it exists, is a small effect.

The equivalence between $f^{\mu}$ and $c_{\mathrm{eff}}^{\nu\mu}$ has a number of interesting consequences---some
relatively straightforward, others rather more subtle. While calculations may be performed 
in either version of the theory, in some cases it may be significantly easier to use one than the other.
It seems that the $c^{\nu\mu}$-type theory is typically simpler to work with, one reason for this being that
effects that appear at first order in $c_{\mathrm{eff}}^{\nu\mu}$ are necessarily second order in $f^{\mu}$.
In fact, from the form of $c_{\mathrm{eff}}^{\nu\mu}$,
it is clear that observable phenomena can only depend on even powers of the $f^{\mu}$
coefficients, and this leads to another curious observation---that the individual components of the $f^{\mu}$
operator do not have physically realizable \textbf{CPT} characteristics. While the operators appear,
according to usual properties of Dirac bilinears, to be odd under \textbf{CPT}, there is no physical
\textbf{CPT} violation, because odd powers of $f^{\mu}$ never appear in observables.

Moreover, because the theories with $f^{\mu}$ and $c^{\nu\mu}_{\mathrm{eff}}$ are entirely equivalent, it seems
that by understanding the quantum corrections in one version of the theory, we should be able to draw conclusions
about the $\mathcal{O}(\hbar)$ behavior of the other.
Taking the $c^{\nu\mu}$ and $f^{\mu}$ terms to be physically equivalent, it would appear that this 
equivalence should manifest itself in the renormalization group
(RG) scalings of the two operator types.
We shall express the RG $\beta$-function for a quantity $x$ using $\beta_{x}=x\Psi(x)$. With this notation,
it seems that at leading order, we ought to have $\Psi(f^{\mu})=\frac{1}{2}\Psi(c^{\nu\rho})$, because this
would make the RG evolution of $c^{\nu\mu}$ and $f^{\nu}f^{\mu}$ the same. However, explicit calculations
of $\beta$-functions did not initially bear this out. Instead, it was found that (regardless of what type of
interactions between fermions were introduced, gauge or Yukawa) $\beta_{f}$ seemed to vanish at leading order,
while $\beta_{c}$ did not.

\begin{figure}[h]
\centering
\includegraphics[angle=0,width=5.4cm]{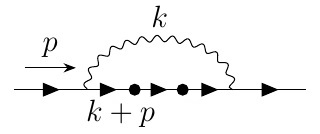}
\caption{Divergent one-loop diagram with two Lorentz-violating $f^{\mu}$ vertex insertions on the internal
fermion line.}
\label{fig-one-loop-f}
\end{figure}

\begin{figure}[h]
\centering
\includegraphics[angle=0,width=2.67cm]{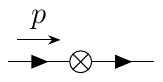}
\caption{Diagram with a single $\delta c^{\nu\mu}$ fermion insertion to cancel the $\mathcal{O}(f^{2})$ divergence
in the diagram from figure~\ref{fig-one-loop-f}.
\label{fig-c-counter}}
\end{figure}

\begin{figure}[h]
\centering
\includegraphics[angle=0,width=12cm]{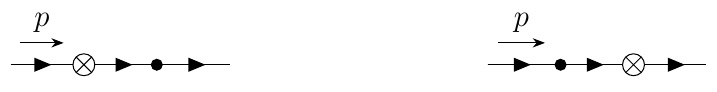}
\caption{The two counterterm diagrams that together cancel the divergence with alternate $f^{\nu}$ and
$\delta f^{\mu}$ insertions.
\label{fig-ff-counter}}
\end{figure}

The reason for this discrepancy was initially somewhat tricky to track down~\cite{ref-karki2}.
It seemed clear that at least part of the difficulty had to be related to the fact that the explicit
calculations of  $\beta_{f}$ were only performed to first order in the Lorentz violation coefficient,
while the equivalence to the $c^{\nu\mu}$ theory appeared at second order. However, it is not necessarily obvious
how this could make a difference in the concrete determination of a $\beta$-function.
The ultimate reason this issue arises
turns out to be that there are different ways that the counterterms may be chosen to
cancel the divergences in the theory with nonzero $f^{\mu}$ coefficients. A one-loop fermion self-energy
diagram  (figure~\ref{fig-one-loop-f}) with two $f^{\mu}$ insertions on the internal fermion line is, as
naively expected, divergent. The divergence can be canceled in either of two different ways: by a single
$\delta c^{\nu\mu}$ counterterm, of the type shown in figure~\ref{fig-c-counter}, or by the pair of diagrams shown
in figure~\ref{fig-ff-counter}, each with two insertions---one of $f^{\nu}$ and one of $\delta f^{\mu}$. Since the
prefactors of the divergent logarithms appearing in the counterterms determine the one-loop $\beta$-functions
for the associated operators, this means that there are two different ways for the SME coefficients to
evolve under changes of the renormalization scale.

In fact, there are more than just two choices for how to renormalize the theory. Linear combinations of the
$\delta c^{\nu\mu}$ diagram and the paired $\delta f^{\mu}$ diagrams may be used to eliminate the divergences.
This means
that there is actually a continuous family of one-loop renormalization schemes, with corresponding continuous
families of $\beta$-function pairs. In a theory in which the force carriers exchanged between fermions
are Yukawa bosons, the divergence in figure~\ref{fig-one-loop-f} may be canceled by any combination of counterterms
\begin{eqnarray}
\delta c^{\nu\mu} & = & -\frac{Xg^{2}}{6}\eta f^{\nu}f^{\mu}
\label{eq-delta-c} \\
\delta f^{\mu} & = & -\frac{(1+2X)g^{2}}{6}\eta f^{\mu},
\end{eqnarray}
where $g$ is the Yukawa coupling, and
\begin{equation}
\eta=\frac{\Gamma(\epsilon/2)}{(4\pi)^{2-\epsilon/2}\mathcal{M}^{\epsilon}}
\end{equation}
is the logarithmic divergence. Any value of the parameter $X$ will work. $X=0$ corresponds to the
choice discussed previously, motivated by the observation that the $f^{\mu}$ is physically equivalent to a
$c^{\nu\mu}$.

Note that if $\delta c^{\nu\mu}$ is nonzero, the theory must actually contain a tree-level $c^{\nu\mu}$\
coefficient as well as an $f^{\mu}$, and there is must be an additional contribution to (\ref{eq-delta-c})
corresponding to the self-renormalization of the $c^{\nu\mu}$ operator. Including the contributions from both
types of SME terms, the RG $\beta$-functions depend on the scheme parameter $X$ according to
\begin{eqnarray}
\label{eq-betacX}
\beta_{c^{\nu\mu}} & = & \frac{2g^{2}}{3(4\pi)^{2}}c^{\nu\mu}
-\frac{Xg^{2}}{3(4\pi)^{2}}f^{\nu}f^{\mu} \\
\label{eq-betafX}
\beta_{f^{\mu}} & = & \frac{(1-X)g^{2}}{3(4\pi)^{2}}f^{\mu}.
\end{eqnarray}
Although equivalence between the $f^{\mu}$ and $c^{\nu\mu}$ coefficients is exact, according
to the first form stated in \eqref{eq-ceff-old}, the specific relations between the
scheme-dependent $\beta$-functions are not
themselves exact, but are valid only in our small $|f^{2}|$ approximation. The scheme dependences of the
RG functions would be more complicated in the exact theory.

In the theory with both $c^{\nu\mu}$ and $f^{\mu}$ appearing in the Lagrangian,
the coefficients cannot actually be measured independently. Up to the order we have considered,
physical quantities such as the fermion energy-momentum relation will only depend on the particular
linear combination $c^{\nu\mu}_{\mathrm{eff}}=c^{\nu\mu}-f^{\nu}f^{\mu}/2$.
According to
(\ref{eq-betacX}--\ref{eq-betafX}), the RG flow for this physically observable combination is given by
\begin{equation}
\label{eq-betacff}
\frac{\partial}{\partial(\log p/\mathcal{M})}
\left(c^{\nu\mu}-\frac{1}{2}f^{\nu}f^{\mu}\right)=\frac{2g^{2}}{3(4\pi)^{2}}
\left(c^{\nu\mu}-\frac{1}{2}f^{\nu}f^{\mu}\right),
\end{equation}
independent of $X$. The physical quantity $c_{\mathrm{eff}}^{\nu\mu}$ has the same RG
evolution as $c^{\nu\mu}$ itself would have in a version of the theory with no $f^{\mu}$ present at all---a
fact which, with a bit of careful consideration, should come as no surprise.

So the RG functions for certain operators in the SME are not unique and instead depend on
the renormalization scheme, although not in a way that seems to lead to any direct inconsistencies.
In fact, it is not unknown for RG $\beta$-functions to depend on the renormalization scheme even in
conventional
Lorentz-invariant QFTs. What is striking here, however, is that the scheme dependence arises already at
one-loop order. Even at the classical level, the description
of the fermion sector contain a redundancy, since it is possible to exchange a $f^{\mu}$ coefficient
for the equivalent $c^{\nu\mu}$ in tree-level calculations. It is not so surprising then that the redundancy
has a manifestation at the quantum level as well.

\section{Discrete Operators in the SME}

Moreover, there is another manifestation of the ambiguity and
the peculiar radiative corrections associated with
it. The Lagrange density $\mathcal{L}_{c}$ is manifestly even under the discrete \textbf{C} and \textbf{CPT}
symmetries. This follows from the usual transformation properties of the vector bilinear in the Dirac theory;
normally, objects with even numbers of Lorentz indices are expected to be \textbf{CPT} even. However, the
equivalent theory described by $\mathcal{L}_{f}$ appears superficially to be \textbf{CPT} odd, because
it involves an operator with a single free index. The necessary resolution is that the physical discrete
\textbf{C}, \textbf{P}, and \textbf{T} operators need to be transformed in accordance with (\ref{eq-f-trans}).
The operators no longer take their usual forms, but in the theory with $f^{\mu}$, they depend explicitly on the
$f^{\mu}$ coefficients, and this creates another interesting situation to explore.

For detailed calculations, we shall focus solely on the parity operator. The reasons for this are
purely pragmatic. The actions of charge conjugation and time reversal involve complex conjugations, and
so the matrix forms of the corresponding operators depend on the representation of the Dirac matrices.
In contrast, the parity operator in the standard theory is always $\gamma^{0}$, independent of how the Dirac
matrices are chosen.

Another comment about the representation of the Dirac matrices is in order at this point. The field
redefinitions that transfer Lorentz violation from the $f^{\mu}$ to $c^{\nu\mu}$ operators may be reinterpreted
as transformations of the Dirac matrices. In essence, a theory in which the only source of Lorentz violation
is a $f^{\mu}$ term is really a theory with a $c^{\nu\mu}$ term and a different choice of representation for
the Clifford algebra of Dirac matrices.

The explicit matrix form of the parity operator in the $\mathcal{L}_{f}$ theory is~\cite{ref-karki5}
\begin{equation}
\textbf{P}\psi\textbf{P}=S_{P}\psi(t,-\vec{x})=
\left[\gamma^{0}+i\frac{f^{0}}{\sqrt{1-f^{2}}}\gamma_{5}
+\frac{\left(1-\sqrt{1-f^{2}}\right)f^{0}f^{\nu}}{f^2\sqrt{1-f^{2}}}\gamma_{\nu}
\right]\psi(t,-\vec{x}).
\label{eq-SP}
\end{equation}
If the $f^{\mu}$ is transformed away according to (\ref{eq-f-trans}), $S_{P}$ transforms into the usual
form $S_{P}'=\gamma^{0}$.
Note also that $S_{P}=\gamma^{0}S_{P}^{\dagger}\gamma^{0}\equiv\bar{S}_{P}$, so the parity
transformation matrix is the same whether it is $S_{P}$ acting to the left of $\psi$ as in (\ref{eq-SP}),
or $\bar{S}_{P}$ acting to the right of $\bar{\psi}$.

Although the form (\ref{eq-SP}) is exact, it may be convenient to take the leading order (in $f^{\mu}$)
approximation, in light of the observed smallness of physical Lorentz violation,
\begin{equation}
\label{eq-SP-approx}
S_{P}\approx\gamma^{0}+if^{0}\gamma_{5}.
\end{equation}
Another simplification occurs if the axial vector $f^{\mu}$ is lightlike. In that case, when the
square roots in (\ref{eq-SP}) are expanded as power series in $f^{2}$, the series terminate with
their leading terms, giving
\begin{equation}
\label{eq-SP-lightlike}
S_{P}=\gamma^{0}+if^{0}\gamma_{5}+\frac{1}{2}f^{0}f^{\nu}\gamma_{\nu}.
\end{equation} 

We can now make more concrete the previous observation that the operators parameterized by
the $f^{\mu}$ coefficients do not typically have well-defined behaviors under discrete spacetime symmetries.
With the modified \textbf{P} operation (\ref{eq-SP}) a spacelike $f^{j}$ term in the Lagrange density
is neither even nor odd under parity.

What is potentially most
consequential about the modified \textbf{P} (and \textbf{C} and \textbf{T}) is that it can
have a nontrivial RG evolution. If the renormalization scheme is chosen with $X\neq 1$, so that
$\beta_{f}\neq 0$, $S_{P}$ will have the $\mathcal{O}(f)$ RG scale dependence
\begin{equation}
\label{eq-SP-RG}
\frac{\partial}{\partial(\log p/\mathcal{M})}S_{P}\approx
i\gamma_{5}\frac{\partial f^{0}}{\partial(\log p/\mathcal{M})}=
i\gamma_{5}\frac{(1-X)g^{2}f^{0}}{3(4\pi)^{2}}.
\end{equation}
So the matrix operator implementing parity is manifestly dependent on the scale unless $X=1$.

For solving the RG equation, the key quantity is the ratio between $\Psi(f^{\mu})$ (governing
the scale evolution of the Lorentz violation coefficient) and $\Psi(g)$ for the
evolution of the Yukawa coupling, in the absence of the SME terms. The conventional
RG behavior of $g$ is
\begin{equation}
\label{eq-g-RG}
g(p)=\frac{g}{\left[1-\frac{10g^{2}}{(4\pi)^{2}}(\log p/\mathcal{M})\right]^{1/2}}.
\end{equation}
The one-loop RG evolution of a SME operator $x$ will be set by the denominator in (\ref{eq-g-RG}),
raised to a power proportional to $\Psi(x)/\Psi(g)$. In particular, with the
$\beta$-function (\ref{eq-betafX}),
\begin{equation}
\label{eq-f-of-p}
f^{\mu}(p)=\frac{f^{\mu}}{\left[1-\frac{10g^{2}}{(4\pi)^{2}}(\log p/\mathcal{M})\right]^{(1-X)/30}}.
\end{equation}
This is sufficient for the solution of the approximate RG equation (\ref{eq-SP-RG}), but
it is actually possible to do quite a bit better.

At this one-loop order, the scaling is the same for all the components of $f^{\mu}$, which means that the
preferred spacetime direction is unchanged by the action of the RG.
Moreover, the fact that all the components of $f^{\mu}(p)$ have the same scale dependence
means that the evolution of
the $f^{\mu}$-exact expression (\ref{eq-SP}) for $S_{P}$ is straightforward. The $\gamma_{5}$ term
inside the brackets in (\ref{eq-SP}) has $f^{0}$ times a power series in $f^{2}$ and thus
contains all the odd powers of $f^{\mu}$ components. The $\gamma_{\nu}$ term contains all the even
powers, with $f^{0}f^{\nu}$ times a different power series in $f^{2}$. The RG dependence of each term
is set simply by the integer degree at which the $f^{\mu}$ components appear in it. For the lightlike case,
the evolution of the terminating expression (\ref{eq-SP-lightlike}) is particularly simple.

\section{Implications}

Whether this observation---that \textbf{C}, \textbf{P}, and \textbf{T} can exhibit nontrivial RG
behavior---has any physically significance may initially seem unclear. Indeed, we may be inclined to think
not, since the RG evolution is explicitly scheme dependent.
However, it is certainly a peculiar observation
meriting more detailed analysis. There are many features of the SME, particularly at the quantum level, that
are not what we might have expected, based solely on experience with conventional Lorentz-invariant
theories, and this is another such a feature.
There is, moreover, at least one somewhat
analogous situation that has definite physical importance---related to the tricky
question of the proton spin and how relatively little of the total nucleon spin (or axial charge)
may appear to derive from the
spins of the underlying quarks. A fermion spin operator may naively appear to be purely a matrix operator
$\Sigma_{0}$
in spinor space, therefore having no scale dependence. However, because of the chiral anomaly, the matrix
elements of the total quark axial charge are modified to~\cite{ref-bass,ref-myhrer}
\begin{equation}
\Sigma=\Sigma_{0}-\frac{N_{f}\alpha_{s}}{2\pi}\Delta G,
\end{equation}
and both the strong coupling $\alpha_{s}$ and the gluon polarization $\Delta G$ are scale-dependent
quantities. While the anomalous scale dependence of the axial charge turned out to be a relatively minor
part of the solution to the ``proton spin crisis,'' it does provide a striking example where the scale
dependence of a seemingly fixed matrix operator has direct physical consequences.

It also turns out that something similar (although not so unequivocal, because of the renormalization
scheme dependence) may be said
about the RG flow of the \textbf{C}, \textbf{P}, and \textbf{T}
operators in the SME with a fermion $f^{\mu}$ term.
While the RG flow of these operators differs from that of $\Sigma$, in that
the scale dependence can always be eliminated by a choice of $X$, the
observation that there are schemes in which these operators have nontrivial RG behavior is nonetheless
important---because there are situations in which considering a theory with $f^{\mu}$ as the only
source of Lorentz violation may actually be quite useful. Any fermion theory with a $c^{\nu\mu}$ tensor that
is a simple tensor product of two four-vectors may be reformulated in terms of a $f^{\mu}$ instead, reducing
the number of SME parameters in the fermion sector from nine to only four.

This reformulation
also simplifies the forms of some important observables in the theory. The impact of $c^{\nu\mu}$-type
Lorentz violation on the behavior of ultrarelativistic fermions is controlled primarily by the
maximum achievable velocity (MAV) of the particles' species. At these high energies, decay and collision
processes are essentially always nearly collinear, because of relativistic beaming. The kinematics and
dynamics look very similar to those of standard $(1+1)$-dimensional special relativity, only
with the uniform speed of light replaced by a direction-dependent MAV, which is (up to first
order in $c^{\nu\mu}$)~\cite{ref-altschul4}
\begin{equation}
v_{\mathrm{MAV}}=1-c_{00}-c_{(0j)}\hat{v}_{j}-c_{jk}\hat{v}_{j}\hat{v}_{k}.
\end{equation}
$\hat{v}$ is a unit three-vector in the direction of the motion, and $c_{(0j)}$
is the symmetrized combination $c_{(0j)}=c_{0j}+c_{j0}$.
If instead of using $c^{\nu\mu}$, the theory is formulated with the corresponding $f^{\mu}$
(satisfying $c^{\nu\mu}\approx-\frac{1}{2}f^{\nu}f^{\mu}$), the expression
for the MAV is more compact and seems to have a clearer physical meaning,
\begin{equation}
v_{\mathrm{MAV}}=1+\frac{1}{2}\left(f^{\mu}\hat{v}_{\mu}\right)^{2},
\end{equation}
where, $\hat{v}^{\mu}$ is now a lightlike four-vector $(1,\hat{v})$.

Owing to the simplicity of this expression, using the equivalent theory with $f^{\mu}$ as the only
source of fermionic Lorentz violation can be extremely advantageous when making measurements
of $v_{\mathrm{MAV}}$---provided the $c^{\nu\mu}$ matrix may be assumed to have the form
of a tensor product. And it actually turns out that $v_{\mathrm{MAV}}$ is frequently the
natural object of study when looking at how high-energy observations can be used to test Lorentz
symmetry. Because of gauge invariance, any charged particle minimally coupled to the electromagnetic field
interacts with photons through its velocity~\cite{ref-altschul17},
and this makes $v_{\mathrm{MAV}}$ a natural
observable in high-energy electron-photon interaction processes. Measurements of the synchrotron and
inverse Compton spectra from high-energy electrons (whether in terrestrial laboratories of
energetic astrophysical environments) may be used to place constraints on $v_{\mathrm{MAV}}$
for the electron field~\cite{ref-altschul7}. Similarly, the observed absences of processes that
are forbidden in conventional relativistic physics, such as vacuum Cerenkov radiation,
$e^{-}\rightarrow e^{-}+\gamma$, or photon decay via $\gamma\rightarrow e^{-}+e^{+}$ also give bounds
on the electron $v_{\mathrm{MAV}}~\cite{ref-stecker,ref-altschul14}$.

So there can certainly be situations when it may be notationally and computationally advantageous
to choose a theory with only $f^{\mu}$-type (as opposed to $c^{\nu\mu}$-type) Lorentz violation.
In such cases, it is also clearly the most convenient to choose a renormalization prescription
with $X=0$, so that if there are no $c^{\nu\mu}$ coefficients at tree level, none will be generated by
one-loop quantum corrections either. In that case, the RG-improved fermion propagator, to
the order represented by the $X=0$ solution (\ref{eq-f-of-p}) is
\begin{equation}
\mathcal{S}_{F}^{(1,f)}(q)=\frac{i}{\slashed{p}+i\!\left[f\!\!\left(\!\!\sqrt{q^{2}}\right)\!\cdot q\right]\!
\gamma_{5}-m},
\label{eq-SF-1f}
\end{equation}
and if we would like to understand the discrete symmetry characteristics of intermediate
states with virtual (off-shell) fermion excitations, it is necessary to use the RG-evolved
\textbf{C}, \textbf{P}, and \textbf{T} operators.

The coefficient of the Lorentz-violating term in the propagator $\mathcal{S}_{F}^{(1,f)}$
depends on $q^{2}$, and this helps with the understanding
of what the scale dependence of $S_{P}$ means. The quantity $q^{2}$ is a measure of how off shell a virtual
particle is. The \textbf{C}, \textbf{P}, and \textbf{T} operations on real fermions, which all have
$q^{2}=m^{2}$, are not affected by the RG flow---which is
reassuring from a physical viewpoint. However, the virtual quanta exchanged
along internal lines in a Feynman diagram behave differently under the discrete transformations
when the actions of \textbf{C}, \textbf{P}, and \textbf{T}
on the entire diagram are considered. For these virtual intermediates, this is a authentic physical effect
if $X\neq1$, not a mathematical artifact.
We may be used to having the action of these operations on real
and virtual Dirac spinors be the same, but there is evidently nothing that requires this to be the case
when Lorentz symmetry is broken.

The standard Feynman propagator has very natural behavior under \textbf{P}. The spatial components
of the four-momentum $q$ are inverted, while the time component is left unchanged,
\begin{equation}
S_{P}\mathcal{S}_{F}^{(0)}S_{P}=S_{P}\!\left(\frac{i}{\slashed{q}-m}\right)\!S_{P}=
\frac{i}{q^{0}\gamma^{0}+\vec{q}\cdot\vec{\gamma}-m}.
\end{equation}
With the inclusion of $c^{\nu\mu}$ coefficients, the generalization is straightforward. The terms in
the denominator of the propagator with $c^{00}$ and $c^{jk}$ are unchanged, but the
$c^{0j}$ term changes sign,
\begin{eqnarray}
S_{P}\mathcal{S}_{F}^{(0,c)}S_{P} & = & S_{P}\!\left(\frac{i}{\slashed{q}+c^{00}q^{0}\gamma^{0}
-c^{0j}q^{j}\gamma^{0}-c^{j0}q^{0}\gamma^{j}+c^{jk}q^{k}\gamma^{j}-m}\right)\!S_{P} \\
& = & \frac{i}{\slashed{q}+c^{00}q^{0}\gamma^{0}
+c^{0j}q^{j}\gamma^{0}+c^{j0}q^{0}\gamma^{j}+c^{jk}q^{k}\gamma^{j}-m}.
\label{eq-c-propagator}
\end{eqnarray}
(The behavior under reflection across a single plane, inverting just one spatial coordinate---in contrast to
\textbf{P}, which inverts all three---is slightly more complicated.)

To get equivalent parity behavior in a $f^{\mu}$ theory that is itself equivalent to a tensor product
$c^{\nu\mu}$ theory, it is necessary to use the modified parity operator of (\ref{eq-SP}) to act on
\begin{equation}
\mathcal{S}_{F}^{(0,f)}=\frac{i}{\slashed{q}+i(f\cdot q)\gamma_{5}-m}.
\end{equation}
With the unmodified parity operation, all four components of the $i(f\cdot q)\gamma_{5}$ in
the denominator would change sign, because $\{\gamma_{5},\gamma^{0}\}=0$. It is necessary to
use the (correct at leading order) $S_{P}=\gamma^{0}+if^{0}\gamma_{5}$ instead to get the proper
result,
\begin{equation}
S_{P}\mathcal{S}_{F}^{(0,f)}S_{P}=\frac{i}{\slashed{q}+if^{0}p^{0}\gamma_{5}
-i\!\left(\!\vec{f}\!\cdot\vec{q}\right)\!\gamma_{5}-m},
\end{equation}
instead. The unchanged $if^{0}q^{0}\gamma_{5}$ component only appears
thanks to the additional term in $S_{P}$ and the way it anticommutes with $\slashed{q}$;
the invariance of this term depends on the $\mathcal{O}(f)$ addition to $S_{P}$ having
the precise coefficient it does.

We can now see that if we instead use the RG-improved propagator $\mathcal{S}_{F}^{(1,f)}$ from
(\ref{eq-SF-1f}) for a virtual fermion intermediate, it will not have the correct parity behavior
(inverting $\vec{q}$ but leaving $q^{0}$ unaffected) unless the parity matrix $S_{P}$
takes the form $S_{P}=\gamma^{0}+if^{0}\!\left(\!\sqrt{q^{2}}\right)\!\gamma_{5}$ at leading order, with the
scale-dependent $f^{0}(p)$ from (\ref{eq-f-of-p}). In other words, in a formulation
of the theory with $f^{\mu}$ coefficients as the only source of fermion-sector Lorentz violation,
it is absolutely necessary to have a parity operator \textbf{P} with nontrivial RG scale
dependence in order to have the correct behavior under discrete symmetries.

Physically, the modifications
to $S_{P}$ must be small. With a coupling constant $g\sim10^{-2}$ and $|f^{\mu}|\lesssim10^{-7}$---as
inferred from the electron $v_{\mathrm{MAV}}$ data---the fractional change in $S_{P}$ with a change of
scale is $\lesssim10^{-14}\,(\log q^{2}/m^{2})$. However, for heavier fermions---especially strongly interacting heavy
quarks---the experimentally established suppression of the RG flow for $S_{P}$ is much less severe,
at a $\lesssim10^{-4}\,(\log q^{2}/m^{2})$ level potentially accessible in precision experiments.
Moreover, regardless
of the experimental status, the potential scale dependence of $S_{P}$ demonstrates the
theoretical possibility of an interesting new feature in QFTs.

The same essential arguments applied to \textbf{P}
hold, rather straightforwardly, for \textbf{C} and \textbf{T}; only keeping track of
additional complex conjugation operators is necessary. We can therefore conclude that there
are definite situations in which the nontrivial RG scalings of \textbf{C}, \textbf{P}, and \textbf{T}
are---if not unavoidable, since we can always choose a formulation of the fermion sector without
a RG-evolving $f^{\mu}$ (or without any $f^{\mu}$ at all)---potentially quite useful and relevant for physically
interesting calculations. This shows again that theories with broken Lorentz symmetry may be interesting
theoretical laboratories for understanding just what kinds of behavior are permitted in QFT.
Especially at the quantum level, these theories have shown interesting types of behavior that are
not normally seen in conventional Lorentz-invariant QFTs.

\section*{Acknowledgments}
The author thanks S. Karki for helpful discussions.

\end{document}